\documentclass[12pt]{article}
\usepackage{latexsym}
\usepackage{amsmath}
\usepackage{amsfonts}
\usepackage{amssymb}
\usepackage[all]{xy}
\usepackage{amscd}
\usepackage{fancybox}
\usepackage{cite}
\def\hybrid{\topmargin -20pt    \oddsidemargin 0pt
        \headheight 0pt \headsep 0pt
        \textwidth 6.25in       
        \textheight 9.5in       
        \marginparwidth .875in
        \parskip 5pt plus 1pt   \jot = 1.5ex}

\hybrid

\newcommand{\beq}{\begin{equation}}
\newcommand{\eeq}{\end{equation}}
\newcommand{\bi}{\begin{itemize}}
\newcommand{\ei}{\end{itemize}}
\newcommand{\bea}{\begin{eqnarray}}
\newcommand{\eea}{\end{eqnarray}}
\newcommand{\ba}{\begin{array}}
\newcommand{\ea}{\end{array}}
\newcommand{\bt}{\begin{tabular}}
\newcommand{\et}{\end{tabular}}
\newcommand{\bc}{\begin{center}}
\newcommand{\ec}{\end{center}}

\def\theequation{\arabic{section}.\arabic{equation}}

\begin{document}

\begin{titlepage}
\begin{center}

\hfill hep-th/0611347\\
\hfill SPIN-06/40\\
\hfill ITP-UU-06/50

\vskip 0.5cm

{\Large \bf Gauged diffeomorphisms and \\hidden symmetries in
Kaluza-Klein theories
\\[0.2cm]}

\vskip 1.5cm

{\bf Olaf Hohm} \\
\vskip 20pt

{\em Spinoza Institute and Institute for Theoretical Physics\\
Leuvenlaan 4, 3584 CE Utrecht,\\
The Netherlands \vskip 5pt }

{email: {\tt o.hohm@phys.uu.nl}} \\

\vskip 0.8cm

\end{center}

\vskip 1cm

\begin{center} {\bf ABSTRACT}\\[3ex]

\begin{minipage}{13cm}
\small We analyze the symmetries that are realized on the massive
Kaluza-Klein modes in generic $D$-dimensional backgrounds with
three non-compact directions. For this we construct the unbroken
phase given by the decompactification limit, in which the higher
Kaluza-Klein modes are massless. The latter admits an
infinite-dimensional extension of the three-dimensional
diffeomorphism group as local symmetry and, moreover, a current
algebra associated to $SL(D-2,\mathbb{R})$ together with the
diffeomorphism algebra of the internal manifold as global
symmetries. It is shown that the `broken phase' can be
reconstructed by gauging a certain subgroup of the global
symmetries. This deforms the three-dimensional diffeomorphisms to
a gauged version, and it is shown that they can be governed by a
Chern-Simons theory, which unifies the spin-2 modes with the
Kaluza-Klein vectors. This provides a reformulation of
$D$-dimensional Einstein gravity, in which the physical degrees of
freedom are described by the scalars of a \textit{gauged}
non-linear $\sigma$-model based on $SL(D-2,\mathbb{R})/SO(D-2)$,
while the metric appears in a purely topological Chern-Simons
form.

\end{minipage}

\end{center}

\noindent

\vfill


November 2006

\end{titlepage}

\section{Introduction} \setcounter{equation}{0}
Kaluza-Klein theories are currently of decisive importance for
modern high-energy physics, not only because of the necessity to
make contact between string- or M-theory and phenomenology, but
also for the conceptual understanding of string theory in general.
For instance, the AdS/CFT correspondence, which provides one of
the rare approaches to non-perturbative effects in string theory,
requires a Kaluza-Klein analysis of the states appearing on
certain AdS backgrounds
\cite{Witten:1998qj,Aharony:1999ti,Skenderis:2006uy}. Second, the
connection between 10-dimensional string theory and 11-dimensional
supergravity or M-theory relies on a Kaluza-Klein reinterpretation
of the spectrum
of D0-branes in type IIA \cite{Polchinski:1998rr}.

Given this importance of Kaluza-Klein theories it is natural to
ask for a detailed understanding of the dynamics of all
Kaluza-Klein modes. However, as the main focus was so far mainly
on phenomenological applications, most approaches have taken only
the lowest or massless modes into account, since the massive modes
are supposed to decouple. In contrast, this is not the case in the
AdS/CFT correspondence. In fact, the internal manifolds appearing
in AdS string backgrounds have to be large \cite{Witten:1998qj},
and therefore the massive modes can no longer be integrated out.
Focusing on the low-energy description, a better understanding of
the effective supergravity actions for massive Kaluza-Klein states
is therefore desirable.

Recently we initiated in \cite{Hohm:2005sc} an analysis of the
effective gravity actions containing the full tower of
Kaluza-Klein modes. Apart from massive states with spin $s\leq 1$
(which, for instance, were incorporated into the effective
supergravity on $AdS_3\times S^3$ in \cite{Nicolai:2003ux}), they
also contain an infinite tower of massive spin-2 states (being the
higher modes of the metric) and, in supergravity, an infinite
tower of massive spin-3/2 fields as their superpartner. A limited
number of massive spin-3/2 states can be described via
spontaneously broken supersymmetry and its associated super-Higgs
effect \cite{Hohm:2005ui,Hohm:2006rj}. Accordingly, the most
natural description of the entire Kaluza-Klein tower would not
only require some infinite-dimensional generalization of local
supersymmetry, but also of the diffeomorphism symmetry.

Based on a circle compactification of pure gravity it has been
argued some time ago in \cite{Dolan:1983aa}, that there is indeed
an infinite-dimensional spontaneously broken gauge symmetry hidden
in the full Kaluza-Klein theory. This infinite-dimensional spin-2
symmetry appears as a remnant of the higher-dimensional
diffeomorphism group. More specifically, every diffeomorphism
generated by a vector field $\xi^M$ gives, upon Fourier expansion,
rise to an infinite-dimensional spin-2 symmetry parametrized by
$\xi^{\mu n}$ (with $n$ denoting the Fourier modes) as well as an
infinite-dimensional gauge symmetry generated by $\xi^{5 n}$. The
latter appears as an ordinary Yang-Mills gauge symmetry, whose Lie
algebra is given by the Virasoro algebra, i.e.~by the
diffeomorphism algebra of the internal manifold (the circle)
\cite{Aulakh:1984qx,Aulakh:1985un,Cho:1991xk,Cho:1992rq,Cho:1992xv}.
The former spin-2 symmetries, on the other hand, have been further
elaborated in \cite{Hohm:2005sc} in the case of a Kaluza-Klein
reduction to $2+1$ dimensions. They are required in order to
guarantee consistency of the gravity--spin-2 couplings in the same
sense that supersymmetry is required for consistency of
gravity--spin-3/2 couplings.

In addition it has been shown that these theories naturally extend
the gauged supergravities in \cite{Nicolai:2000sc,Nicolai:2001sv}
in the sense that they are also deformations of an unbroken phase
with an enhanced global symmetry, while the gauge fields only
enter through Chern-Simons terms. This unbroken phase corresponds
to the decompactification limit and can formally be determined
simply be restoring in the zero-mode action the dependence of all
fields on the internal coordinate. In more mathematical terms this
means to replace the metric and all matter fields by an
`algebra-valued' object, where the (commutative and associative)
algebra is given by the algebra of smooth functions on the circle.
This fits into a non-standard form of general relativity
introduced by Wald \cite{Cutler:1986dv,Wald:1986dw}, which is
based on a so-called algebra-valued differential geometry, and
which is essentially the only way to get a multi-graviton theory
that is consistent with a generalized diffeomorphism or spin-2
symmetry \cite{Aragone:1979bm,Boulanger:2000rq,Duff:1989ea}.
Furthermore, upon performing duality transformations in $D=3$ it
has been shown that the hidden symmetry realized on the zero-modes
(the Ehlers group $SL(2,\mathbb{R})$) gets enhanced to its affine
extension. The full (massive) Kaluza-Klein theory can then be
reconstructed by gauging a certain subalgebra of this rigid
symmetry, i.e.~by promoting this subalgebra to a local symmetry.
This in turn deforms the diffeomorphisms in the sense that, in
particular, each partial derivative in the standard formulas for
diffeomorphisms gets replaced by a gauge-covariant derivative
$D_{\mu}$. Due to the non-commutativity $[D_{\mu},D_{\nu}]\sim
F_{\mu\nu}$ this turns to a symmetry which is no longer manifest.
However, in the (2+1)-dimensional context the spin-2 fields and
the gauge vectors combine into a Chern-Simons theory
\cite{Witten:1988hc,Achucarro:1987vz,Blencowe:1988gj} based on a
Lie algebra which contains a non-standard semi-direct product
between the Virasoro algebra and a centrally extended Kac-Moody
algebra associated to the Poincar\'e group. This allows a direct
investigation of the gauged diffeomorphisms, due to which the
compactification to $D=3$ is an excellent arena for the analysis
of these symmetries.

One aim of the present paper is to show that these results extend
to more general Kaluza-Klein backgrounds, which again compactify
to $D=3$,\footnote{Kaluza-Klein compactifications to $D=4$ without
truncation have been considered in \cite{Aulakh:1985my}.} but
whose internal manifolds can be more complex. We will see that the
existence of a Chern-Simons description as well as an enhancement
of the global symmetries is a generic feature, but that the
associated symmetry algebras get more involved: The Virasoro
algebra and the (affine) Kac-Moody algebras appearing in $S^1$
compactifications are replaced by the diffeomorphism algebra of
the internal manifold (whose form we will give explicitly in case
of a torus) and the Lie algebra of so-called current groups. The
latter generalize the affine algebras as the Lie algebras of a
Loop group $C^{\infty}(S^1,G)$ associated to a Lie group $G$ to
$C^{\infty}(K,G)$ for arbitrary compact manifolds $K$. These are
substantially more intricate than Loop groups and so have not been
studied exhaustively in the mathematical literature
\cite{Pressley:1988qk,Mickelsson:1989hp}. Investigations of
Kaluza-Klein theories might therefore also be of interest in this
respect.

A second motivation for the present analysis is to give a
reformulation of $D$-dimensional Einstein gravity in a form that
may shed light on the role that the so-called `hidden symmetries'
encountered in dimensional reductions play in the original theory
\cite{Cremmer:1997ct,Cremmer:1999du,Lambert:2001gk}. In fact, once
all massive Kaluza-Klein modes are taken into account, the
Kaluza-Klein theory can still be viewed as being $D$-dimensional,
but in a particular -- Kaulza-Klein inspired -- Lorentz gauge.
This approach has been pioneered in
\cite{deWit:1986mz,Nicolai:1986jk}, where it has been shown that
part of the hidden symmetries appearing via reducing
11-dimensional supergravity to $D=4$ and $D=3$, respectively, can
be seen already in 11 dimensions, upon fixing part of the Lorentz
symmetry. More specifically, the composite local symmetry groups
appearing in the coset spaces $E_{7(7)}/SU(8)$ and
$E_{8(8)}/SO(16)$ exist also in $11$-dimensional supergravity,
while the role of the exceptional groups remains somewhat
mysterious. (See, however, \cite{Koepsell:2000xg}.) In a similar
spirit it has been suggested that gravity in $D$ dimensions should
have an interpretation as a non-linear $\sigma$-model based on
$SL(D-2,\mathbb{R})/SO(D-2)$, which is exactly the structure that
appears by reducing to $D=3$ on a torus. It might therefore be of
interest that, as we are going to show in this paper, gravity in
any dimension can be seen as a \textit{gauged} non-linear
$\sigma$-model of this type (in a sense that we will make precise
below). More specifically, the ungauged theory, still being fully
$D$-dimensional, admits the entire $SL(D-2,\mathbb{R})$ as
symmetry group, whose breaking in the full theory is due to the
gauging. This should be compared with \cite{Goroff:1983hc}, in
which $SL(D-2,\mathbb{R})$ has been realized as a symmetry of
$D$-dimensional gravity in light-cone gauge, but with the action
in a non-local form.

The paper is organized as follows. After briefly reviewing the
structure of Kaluza-Klein theories and symmetries for
compactifications to $D=3$ in sec.~2, we discuss in sec.~3 the
unbroken phase together with its symmetries. In sec.~4 we turn to
the problem of reconstructing the full $D$-dimensional
gravitational theory, i.e.~the `broken phase', via gauging certain
symmetries. The appearing consistency problems are discussed, and
it is shown that their resolution can be made manifest within a
particular subsector given by a Chern-Simons description. We
conclude in sec.~5. Appendix A reviews the emergence of the hidden
symmetry $SL(D-2,\mathbb{R})$ in torus reductions to $D=3$, while
appendix B shows the details of an explicit Kaluza-Klein analysis
without truncation.

\section{Kaluza-Klein theory on $\mathbb{R}^{1,2}\times K_d$}
\setcounter{equation}{0} Here we give a brief review of
Kaluza-Klein theories on backgrounds of the form
$\mathbb{R}^{1,2}\times K_d$. At this stage $K_d$ has to be
Ricci-flat, since we are dealing with pure Einstein gravity, even
though we will later see that this assumption can be relaxed.
Whenever we consider the action and symmetry transformations in
terms of a mode expansion, we will specialize to a $d$-dimensional
torus. This should, however, not be confused with a particular
truncation, since we will keep the dependence on all $D=3+d$
coordinates.

Our starting point is pure gravity in $D$ dimensions, described by
the Einstein-Hilbert action
 \bea
  S_{\rm EH}=-\int d^{3+d}x\hspace{0.1cm}ER\;.
 \eea
We make a Kaluza-Klein ansatz by fixing the Lorentz symmetry, so
that the vielbein appears in a triangular gauge:\footnote{Our
conventions are as follows: The coordinates are
$x^M=(x^{\mu},\hat{y}^{m})\equiv (x^{\mu},g^{-1}y^m)$. Space-time
and Lorentz indices are labelled in $D$ dimensions by $M,N,K,...$
and $A,B,C,...$, in $2+1$ dimensions by $\mu,\nu,\rho,...$ and
$a,b,c,...$, and finally for the internal $d$ dimensions by
$m,n,k,...$ and $\alpha,\beta,\gamma,...$, respectively. The
metrics are mostly minus.}
 \bea\label{vielbein}
  E_M^A=\left(\begin{array}{cc} \phi^{-1}e_{\mu}^a &
  A_{\mu}^m\phi_m^{\alpha} \\ 0 & \phi_m^{\alpha} \end{array}\right)\;.
 \eea
Here $\phi_m^{\alpha}$ are scalar fields, of which we may think as
parametrizing the vielbein of the internal manifold, and
 \bea
   \phi=\det (\phi_m^{\alpha})=\frac{1}{d!}\epsilon^{m_1 ... m_d}
   \epsilon_{\alpha_1 ... \alpha_d}
   \phi_{m_1}^{\alpha_1}...\phi^{\alpha_d}_{m_d}\;.
 \eea
The inverse vielbein reads
 \bea
  E_A^M=\left(\begin{array}{cc} \phi e_a^{\mu} &
  -e_a^{\rho}A_{\rho}^m\phi \\ 0 & \phi_{\alpha}^m \end{array}\right)\;,
 \eea
where $\phi_{\alpha}^m$ denotes the inverse of $\phi_m^{\alpha}$.

The dimensionally reduced action or, equivalently, the zero-mode
action in case of a torus, takes the following form
 \bea\label{KKaction}
 \begin{split}
  S=\int d^3x\hspace{0.2em} e
  \big [ &-R^3(e)-\frac{1}{4}\phi^2G_{mn}(\phi)
  F^{\mu\nu\hspace{0.1em}m}F_{\mu\nu}^n+\phi^{-2}\partial^{\mu}\phi
  \partial_{\mu}\phi \\
  &+\frac{1}{2}g^{\mu\nu}(\phi_{\alpha}^m\partial_{\mu}\phi_m^{\gamma})
  (\phi_{\gamma}^n\partial_{\nu}\phi_n^{\alpha})-\frac{1}{2}G^{mn}(\phi)
  g^{\mu\nu}\partial_{\mu}\phi_m^{\beta}\partial_{\nu}\phi_{n\beta}\big]\;.
 \end{split}
 \eea
Here the gauge kinetic couplings are defined by
$G_{mn}=\delta_{\alpha\beta}\phi_m^{\alpha}\phi_n^{\beta}$. After
this truncation the only remnant of the $D$-dimensional
diffeomorphisms are $3$-dimensional diffeomorphisms and $U(1)^d$
gauge tranformations, for which
$F_{\mu\nu}^m=\partial_{\mu}A_{\nu}^m-\partial_{\nu}A_{\mu}^m$
provides the invariant field strength.

Let us next analyze which form the full $D$-dimensional
diffeomorphisms take in the Kaluza-Klein gauge (\ref{vielbein}),
or in other words, which symmetry is realized on the full tower of
Kaluza-Klein modes without truncation. The $D$-dimensional
diffeomorphisms and local Lorentz transformations are parametrized
by $\xi^M$ and $\Lambda^A_{\hspace{0.3em}B}$, respectively, and
read
 \bea
  \delta_{\xi}E_M^A = \xi^N\partial_N E_M^A + \partial_M\xi^N
  E_N^A\;, \qquad
  \delta_{\Lambda}E_M^A=\Lambda^A_{\hspace{0.3em}B}E_M^B\;.
 \eea
Splitting the diffeomorphisms as $\xi^M=(\xi^{\mu},\xi^m)$, they
act on the Kaluza-Klein fields according to
 \bea\label{diffaction}
 \begin{split}
  \delta_{\xi}\phi_m^{\alpha}&=\xi^{\rho}\partial_{\rho}\phi_m^{\alpha}
  +g\xi^n\partial_n\phi_m^{\alpha}+g\partial_m\xi^{\rho}A_{\rho}^n
  \phi_n^{\alpha}+g\partial_m\xi^n\phi_n^{\alpha}\;, \\
  \delta_{\xi}\phi&=\xi^{\rho}\partial_{\rho}\phi+g\xi^n\partial_n\phi
  +g\partial_m\xi^{\rho}A_{\rho}^m\phi+g\partial_m\xi^m\phi\;, \\
  \delta_{\xi}A_{\mu}^m&=\xi^{\rho}\partial_{\rho}A_{\mu}^m
  +g\xi^n\partial_nA_{\mu}^m+\partial_{\mu}\xi^{\rho}A_{\rho}^m
  +\partial_{\mu}\xi^m-gA_{\mu}^n\partial_n\xi^{\rho}A_{\rho}^m
  -gA_{\mu}^n\partial_n\xi^m\;, \\
  \delta_{\xi}e_{\mu}^a&=\xi^{\rho}\partial_{\rho}e_{\mu}^a
  +g\xi^m\partial_me_{\mu}^a+\partial_{\mu}\xi^{\rho}e_{\rho}^a
  +gA_{\rho}^m\partial_m\xi^{\rho}e_{\mu}^a+g\partial_m\xi^me_{\mu}^a\;.
 \end{split}
 \eea
Here we have introduced a parameter $g$, which will later on serve
as gauge coupling constant. In the case of a torus it corresponds
to the radii $R=g^{-1}$, which are taken to be equal.
The transformations (\ref{diffaction}) are in general not
compatible with the triangular gauge in (\ref{vielbein}). Thus we
have to add a compensating Lorentz transformation, which turns out
to be given by
 \bea
  \Lambda^a_{\hspace{0.3em}\alpha}=-g\phi^{-1}\phi_{\alpha}^{\hspace{0.2em}m}
  \partial_m\xi^{\rho}e_{\rho}^a\;.
 \eea
This yields
 \bea\label{complor}
  \delta_{\Lambda}\phi_m^{\alpha}=0\;, \quad
  \delta_{\Lambda}e_{\mu}^a=-gA_{\mu}^m\partial_m\xi^{\rho}e_{\rho}^a\;, \quad
  \delta_{\Lambda}A_{\mu}^m=-g\phi^{-2}G^{mn}\partial_n\xi^{\rho}
  g_{\rho\mu}\;.
 \eea

Next one can perform a mode expansion associated to a torus, which
reads for the scalar, e.g.,
 \bea\label{fourier}
  \phi(x^{\mu},y^m)=\sum_{n_1=-\infty}^{\infty} ...
  \sum_{n_d=-\infty}^{\infty} \phi^{[n_1,...,n_d]}(x)e^{i n_1 y^1} ...
  \hspace{0.4em}e^{in_d y^d}\;,
 \eea
and similarly for all other fields. Moreover, we have to impose a
reality constraint on the fields, as
$(\phi^{[n_1,...,n_d]})^*=\phi^{[-n_1,...,-n_d]}$. Also the
transformation parameter can be expanded into Fourier modes. This
results then in an infinite-dimensional symmetry, which is
spontaneously broken to the symmetry of the zero-modes
\cite{Dolan:1983aa}. The global symmetry and its Lie algebra in
the unbroken phase will be determined in the next section.

\section{The unbroken phase and its symmetries}\label{rigid}
As explained in the introduction, the unbroken phase, in which the
spin-2 symmetries are manifestly and linearly realized, can
essentially be reconstructed simply by restoring the dependence on
the internal coordinates $y^m$ of all fields in the dimensionally
reduced action (\ref{KKaction}). This results in
 \bea\label{ungaugedaction}
  \begin{split}
   S_{\rm 0}=\int d^3x d^d y\hspace{0.2em} e
   \big [ &-R^3(e)-\frac{1}{4}\phi^2G_{mn}(\phi)
   F^{\mu\nu\hspace{0.1em}m}F_{\mu\nu}^n+\phi^{-2}\partial^{\mu}\phi
   \partial_{\mu}\phi \\
   &+\frac{1}{2}g^{\mu\nu}(\phi_{\alpha}^m\partial_{\mu}\phi_m^{\gamma})
   (\phi_{\gamma}^n\partial_{\nu}\phi_n^{\alpha})-\frac{1}{2}G^{mn}(\phi)
   g^{\mu\nu}\partial_{\mu}\phi_m^{\beta}\partial_{\nu}\phi_{n\beta}\big]\;.
  \end{split}
 \eea
As in contrast to (\ref{KKaction}) the fields depend on all $D$
coordinates the action contains also an integration over the
additional $d$ internal coordinates. In this sense
(\ref{ungaugedaction}) describes a truly $D$-dimensional theory,
but without the full $D$-dimensional diffeomorphism and Lorentz
invariance. As we are going to show in the following, this theory
might instead be viewed as the Kaluza-Klein theory in the
decompactification limit $R\rightarrow \infty$ (i.e.~$g\rightarrow
0$), with all Kaluza-Klein modes retained.

It can in turn be seen that, compared to (\ref{KKaction}), in
(\ref{ungaugedaction}) an enhancement of the three-dimensional
diffeomorphism symmetry takes place. More precisely, with the
standard formulas for diffeomorphisms it can be easily checked
that they leave (\ref{ungaugedaction}) invariant also if the
transformation parameter $\xi^{\mu}$ is allowed to depend on the
internal coordinates. Explicitly, they act on the fields as
(\ref{diffaction}) with $\xi^m=0$, $g=0$.

Besides this local infinite-dimensional diffeomorphism or spin-2
symmetry, there appears also an infinite-dimensional global
symmetry group. The latter is the rigid remnant of the
diffeomorphism group of the internal manifold. They act like the
$\xi^m$-variations in (\ref{diffaction}), but with $\xi^m$ being
independent of space-time, and it can be easily checked that they
leave (\ref{ungaugedaction}) invariant. In contrast to the
compactification on a circle \cite{Hohm:2005sc}, this group is no
longer defined by the Virasoro or Witt algebra, but instead by a
more complicated algebra. More specifically, in case that the
internal manifold is a torus, the algebra, which we will denote in
the following by $\hat{v}_d$, is spanned by generators
$Q^{m[j_1,...,j_d]}$ and reads
 \bea
  [Q^{m[j_1,...,j_d]},Q^{n[k_1,...,k_d]}]=
  i\left(j_nQ^{m[j_1+k_1,...,j_d+k_d]}-k_mQ^{n[j_1+k_1,...,j_d+k_d]}\right)\;.
 \eea
The subalgebra spanned by all generators of the form
$Q^m_j:=Q^{j[m,...,m]}$, where $j=1,...,d$, takes the form
 \bea
  [Q_j^m,Q_j^n]=i(m-n)Q_j^{m+n}\;,
 \eea
and thus the algebra contains, as expected, $d$ copies of the
Virasoro algebra. Note, however, that it is not a direct sum,
since the $Q^m_j$ do not commute for different $j$.

As one of the results of \cite{Hohm:2005sc} it has been found that
upon dualization the theory can equivalently be written in a form
that admits moreover the affine extension of the hidden symmetry
group $SL(2,\mathbb{R})$ (the Ehlers group) as global symmetry
group. In general, for reductions on $d$-dimensional tori to $D=3$
a hidden $SL(d+1,\mathbb{R})$ symmetry appears. (See appendix A
for a review.) Thus one might expect that in the unbroken phase
also the latter extends to a symmetry on the full Kaluza-Klein
tower and has moreover an infinite-dimensional extension. In the
following we are going to show that this is indeed the case.

The corresponding action can equally be determined from the
zero-mode action in the form (\ref{dualaction}), where all degrees
of freedom have been dualized into scalars:
 \bea\label{ungaugeddual}
  \begin{split}
   S_0=\int d^3x d^d y \hspace{0.2em}e\big[-R^3(e)&+g^{\mu\nu}(\phi^{-2}
   \partial_{\mu}\phi\partial_{\nu}\phi+\frac{1}{2}\phi^{-2}G^{mn}(\phi)
   \partial_{\mu}\varphi_m\partial_{\nu}\varphi_n \\
   &+\frac{1}{2}\phi_{\alpha}^m\partial_{\mu}\phi_m^{\gamma}\phi_{\gamma}^n
   \partial_{\nu}\phi_n^{\alpha}-\frac{1}{2}G^{mn}(\phi)\partial_{\mu}
   \phi_m^{\beta}\partial_{\nu}\phi_{n\beta})\big]\;.
  \end{split}
 \eea
Here the dual scalars $\varphi_m$ are defined by means of the
duality relation (\ref{dualrel}) given in appendix A, but again
with all fields depending also on the $y^m$. The
$SL(d+1,\mathbb{R})$ symmetry transformations given in appendix A
can then depend also on the internal coordinates without affecting
the invariance of the action. The underlying Lie algebra is,
however, more complicated than the affine, that is, loop group
extension which appears in case of an $S^1$ compactification.
Instead the group is given by the smooth maps from the compact
manifold $K_d$ into the considered Lie group,
$C^{\infty}(K_d,SL(d+1,\mathbb{R}))$. In the mathematical
literature these are known as current groups. In case that the
compact manifold is given by the torus $T^d$ its Lie algebra will
be denoted in the following by $T^d
sl(d+1,\mathbb{R})$.\footnote{In the case $d=1$ this reduces to
the affine extension of $SL(2,\mathbb{R})$ discussed in
\cite{Hohm:2005sc}.} Like in the case of affine Kac-Moody
algebras, the algebra $T^d\frak{g}$ associated to any
finite-dimensional Lie algebra $\frak{g}$ can essentially be
determined by endowing the generators of $\frak{g}$ with the
exponential Fourier modes as in (\ref{fourier}). Let us illustrate
this for $sl(d+1,\mathbb{R})$. In the basis defined in appendix A
the generators are given by $K_{\hspace{0.3em}b}^{a}$ ($a,b =
1,...,d$), which span the $sl(d,\mathbb{R})$ subalgebra, as well
as $e_a$, $f^a$ and $\hat{e}$, whose transformation properties
under $sl(d,\mathbb{R})$ are given in (\ref{liealgebra}). The
current subalgebra $T^d sl(d,\mathbb{R})$ can then simply be read
off from (\ref{sld}),
 \bea
  [K_{\hspace{0.3em}b}^{a[j_1,...,j_d]},K_{\hspace{0.3em}d}^{c[k_1,...,k_d]}]
  = \delta_d^a K_{\hspace{0.3em}b}^{c[j_1+k_1,...,j_d+k_d]}-\delta_b^c
  K_{\hspace{0.3em}d}^{a[j_1+k_1,...,j_d+k_d]}\;,
 \eea
while for the remaining brackets one finds from (\ref{liealgebra})
 \bea
  \begin{split}
   [K_{\hspace{0.3em}b}^{a[\underline{j}]},e_c^{[\underline{k}]}] &= \delta_c^a e_b^{[\underline{j}+\underline{k}]}
   - \frac{1}{d}\delta_b^a e_c^{[\underline{j}+\underline{k}]}\;,
   \qquad
   [K_{\hspace{0.3em}b}^{a[\underline{j}]},f^{c[\underline{k}]}] = -\delta^c_b f^{a[\underline{j}+\underline{k}]}
   + \frac{1}{d}\delta^a_b f^{c[\underline{j}+\underline{k}]};,\\
   [e_a^{[\underline{j}]},f^{b[\underline{k}]}] &= K_{\hspace{0.3em}a}^{b[\underline{j}+\underline{k}]}
   - \frac{1}{d}\hat{e}^{[\underline{j}+\underline{k}]}\delta_a^b\;, \quad
   [K_{\hspace{0.3em}b}^{a[\underline{j}]},\hat{e}^{[\underline{k}]}] = 0\;. \\
   [e_a^{[\underline{j}]},\hat{e}^{[\underline{k}]}] &= (d+1)e_a^{[\underline{j}+\underline{k}]} \;, \qquad
   [f^{a[\underline{j}]},\hat{e}^{[\underline{k}]}] = -(d+1)f^{a[\underline{j}+\underline{k}]}\;, \\
   [e_a^{[\underline{j}]},e_b^{[\underline{k}]}]&=0\;, \qquad [f_a^{[\underline{j}]},f_b^{[\underline{k}]}]=0\;.
  \end{split}
 \eea
Here we have introduced the compact notation
$Q^{m[\underline{j}]}$, etc., where $[\underline{j}]$ denotes the
row vector $[j_1,...,j_d]$.

To find the full rigid symmetry algebra in the unbroken phase, we
have to ask if there is some non-trivial product between the
diffeomorphism algebra $\hat{v}_d$ and the current algebra
associated to $SL(d+1,\mathbb{R})$. In the case of a circle
compactification this reduces -- upon a simple change of basis --
to the standard form of a semi-direct product between a Virasoro
algebra and a Kac-Moody algebra \cite{Hohm:2005sc}, which is
well-known to physicists from the Sugawara construction. In the
more general case, however, the resulting structure is not clear a
priori and therefore will be analyzed in the following.

First of all, we have to know the action of the internal
diffeomorphism algebra $\hat{v}_d$ on all physical fields, in
particular on the dual scalar $\varphi_m$ (carrying the former
degrees of freedom of the Kaluza-Klein vectors $A_{\mu}^m$). This
can be determined by applying the $\xi^m$ variations in
(\ref{diffaction}) to the duality relation (\ref{dualrel}). One
finds after some computations\footnote{Here we have rescaled the
transformation parameter such that in the ungauged phase no factor
of $g$ appears.}
 \bea
  \delta_{\xi}\varphi_m = \xi^k\partial_k\varphi_m + \partial_k\xi^k
  \varphi_m + \partial_m\xi^k\varphi_k\;.
 \eea
In order to determine the `semi-direct' product let us first check
the closure of the symmetry variations (with $y^m$-dependent
parameter). One finds
 \bea\label{semi}
  \begin{split}
   [\delta_{\xi}(Q),\delta_{\lambda}(K)]\phi_m^{\alpha} &=
   \delta_{\tilde{\lambda}}(K)\phi_m^{\alpha}\;, \\
   [\delta_{\xi}(Q),\delta_{\kappa}(e)]\varphi_m &=
   \delta_{\tilde{\kappa}}(e)\varphi_m\;, \\
   [\delta_{\xi}(Q),\delta_{\sigma}(f)]\phi_m^{\alpha} &=
   \delta_{\tilde{\sigma}}(f)\phi_m^{\alpha}\;, \\
   [\delta_{\xi}(Q),\delta_{\epsilon}(\hat{e})]\phi_m^{\alpha}
   &=\delta_{\tilde{\epsilon}}(\hat{e})\phi_m^{\alpha}\;,
  \end{split}
 \eea
where the transformation parameter are given by
 \bea\label{transparam}
  \begin{split}
   \tilde{\lambda}_{\hspace{0.3em}m}^k
   &= -\xi^n\partial_n\lambda_{\hspace{0.3em}m}^k
   +\lambda_{\hspace{0.3em}m}^n\partial_n\xi^k - \partial_m\xi^n
   \lambda_{\hspace{0.3em}n}^k\;, \\
   \tilde{\kappa}_m &=
   -\xi^k\partial_k\kappa_m-\partial_k\xi^k\kappa_m-\partial_m\xi^k\kappa_k\;,
   \\
   \tilde{\sigma}^m
   &=-\xi^k\partial_k\sigma^m+\partial_k\xi^k\sigma^m+\sigma^k\partial_k\xi^m\;,
   \\
   \tilde{\epsilon} &=-\xi^k\partial_k \epsilon\;.
  \end{split}
 \eea
By expanding (\ref{transparam}) into Fourier modes one can read
off the Lie algebra:
 \bea\label{semialg}
  \begin{split}
   [Q^{m[\underline{j}]},K^{n[\underline{k}]}_{\hspace{0.3em}p}] &=
   -ik_mK^{n[\underline{j}+\underline{k}]}_{\hspace{0.3em}p}
   +ij_pK^{n[\underline{j}+\underline{k}]}
   _{\hspace{0.3em}m}-ij_q\delta^{mn} K^{q[\underline{j}+\underline{k}]}
   _{\hspace{0.3em}p}\;, \\
   [Q^{p[\underline{j}]},f_q^{[\underline{k}]}] &=
   i\big(-k_pf_q^{[\underline{j}+\underline{k}]}+j_pf_q^{[\underline{j}
   +\underline{k}]}+j_qf_p^{[\underline{j}+\underline{k}]}\big)\;, \\
   [Q^{p[\underline{j}]},e^{q[\underline{k}]}] &=
   -i\left((k_p+j_p)e^{q[\underline{j}+\underline{k}]}+\delta^{pq}j_l
   (e^l)^{[\underline{j}+\underline{k}]}\right)\;, \\
   [Q^{n[\underline{j}]},\hat{e}^{[\underline{k}]}] &=
   -ik_n \hat{e}^{[\underline{j}+\underline{k}]}\;.
  \end{split}
 \eea
Note that this Lie algebra reduces for $d=1$ to the algebra of
\cite{Hohm:2005sc}, if one identifies (in the notation of
\cite{Hohm:2005sc}) $h$ with $\hat{e}$, $f$ with $f_1$ and $e$
with $e^1$, while $K_{\hspace{0.3em}b}^a$ trivializes (in order to
be traceless).

One can check explicitly that this defines a consistent Lie
algebra satisfying the Jacobi identities, generalizing the
well-known semi-direct product between the standard Virasoro
algebra and an affine Kac-Moody algebra.

\section{Reconstruction of $D$-dimensional Einstein gravity}
In the last section we have shown that the dual action
(\ref{ungaugeddual}) is invariant under an infinite-dimensional
diffeomorphism symmetry and an infinite-dimensional extension of
the hidden rigid invariance group $SL(d+1,\mathbb{R})$. It is
remarkable that a truly $D$-dimensional action still admits the
$SL(D-2,\mathbb{R})$ symmetry. However, the diffeomorphism group
of the internal manifold is also realized only as a global
symmetry on the full Kaluza-Klein tower, or equivalently, locally
only in the internal coordinates.

As a next step we are going to reconstruct the full
$D$-dimensional gravity theory. First we have to promote the
internal diffeomorphism algebra to a local symmetry, i.e. we have
to gauge the subalgebra $\hat{v}_d$. In the next subsection we are
going to determine the covariant derivatives, which are required
for a minimal coupling, and we will briefly discuss the appearing
consistency problems related to `gauged diffeomorphisms'. Then we
discuss the manifest resolution of these consistency problems for
a particular subsector of the theory, namely the sector consisting
of gravitational and gauge fields. Finally we give the full
action, which is on-shell equivalent to the original
$D$-dimensional gravity theory.

\subsection{Covariantisation and gauged
diffeomorphisms}\label{covsection} To begin with, we have to
replace all partial derivatives by covariant derivatives with
respect to $\hat{v}_d$. These can be most conveniently written by
adding additional terms proportional to the internal derivative
$\partial_m$. They read
 \bea\label{covder}
  \begin{split}
   D_{\mu}\phi_m^{\alpha}&= \partial_{\mu}\phi_m^{\alpha}
   -gA_{\mu}^n\partial_n\phi_m^{\alpha}-g\phi_n^{\alpha}\partial_mA_{\mu}^n\;, \\
   D_{\mu}\phi &= \partial_{\mu}\phi-gA_{\mu}^n\partial_n\phi
   -g\phi\partial_nA_{\mu}^n\;, \\
   D_{\mu}\varphi_m &= \partial_{\mu}\varphi_m -g A_{\mu}^n\partial_n\varphi_m
   -g\varphi_n\partial_m A_{\mu}^n - g\varphi_m\partial_kA_{\mu}^k\;, \\
   D_{\mu}e_{\nu}^a &=\partial_{\mu}e_{\nu}^a-gA_{\mu}^m\partial_m
   e_{\nu}^a-ge_{\nu}^a\partial_mA_{\mu}^m\;, \\
   D_{\mu}\omega_{\nu}^a &= \partial_{\mu}\omega_{\nu}^a-gA_{\mu}^m
   \partial_m\omega_{\nu}^a\;.
  \end{split}
 \eea
Here $\omega_{\mu}^a$ denotes the spin connection, which
transforms under $\hat{v}_d$ as
 \bea
  \delta_{\xi}\omega_{\mu}^a = g\xi^m\partial_m\omega_{\mu}^a\;.
 \eea
One can easily check that the covariant derivatives (\ref{covder})
transform covariantly under local $\hat{v}_d$ transformations,
e.g.
 \bea
  \delta_{\xi^m}(D_{\mu}\phi_m^{\alpha}) =
  g\xi^n\partial_n(D_{\mu}\phi_m^{\alpha}) + g\partial_m\xi^n
  D_{\mu}\phi_n^{\alpha}\;.
 \eea
Comparing with (\ref{diffaction}) we see that
$D_{\mu}\phi_m^{\alpha}$ transforms exactly as $\phi_m^{\alpha}$
and similarly for all other fields. Thus, at this stage
$\hat{v}_d$ is manifestly realized.

Apart from the problem that we did not yet introduce a kinetic
term for the $\hat{v}_d$ gauge fields $A_{\mu}^m$, we have to ask
the question if the spin-2 transformations or generalized
diffeomorphisms parametrized by $\xi^{\mu}$ are still a symmetry.
This is not the case, simply due to the fact that the internal
derivatives $\partial_m$ in the covariant derivatives also act on
the spin-2 transformation parameter $\xi^{\rho}$. Put differently,
the covariant derivative of, say, a scalar, will not transform
like a 1-form, but will pick up a non-covariant piece. In order to
keep spin-2 invariance we also have to deform the diffeomorphisms
by $g$-dependent terms. Their actual form can partially be
determined by requiring closure of local $\hat{v}_d$ with spin-2
transformations. For the scalars this implies
 \bea
  \delta_{\xi}\phi_{m}^{\alpha}=\xi^{\rho}\partial_{\rho}\phi_m^{\alpha}
  +g\phi_n^{\alpha}\partial_m\xi^{\rho}A_{\rho}^n\;,
 \eea
since then the algebra closes according to
 \bea\label{closure}
  [\delta_{\xi},\delta_{\eta}(Q)]\phi_m^{\alpha}=\delta_{\tilde{\xi}}\phi_m^{\alpha}
  +\delta_{\tilde{\eta}}(Q)\phi_m^{\alpha}\;,
 \eea
where the transformation parameter are given by
 \bea\label{cloparam}
  \tilde{\xi}^{\rho}=\eta^n\partial_n\xi^{\rho}\;, \qquad
  \tilde{\eta}^n = -\xi^{\rho}\partial_{\rho}\eta^n\;.
 \eea
Comparing with (\ref{diffaction}) one infers that requiring
closure of the symmetry variations allows to recover the expected
spin-2 transformations for $\phi_m^{\alpha}$. Similarly, one shows
that for $e_{\mu}^a$ the spin-2 variations get deformed in the
gauged phase according to (\ref{diffaction}). However, the
variations for $A_{\mu}^m$ determined like this do not coincide
completely with (\ref{diffaction}) and (\ref{complor}), since the
term depending on $\phi$ is not necessary for the closure as in
(\ref{closure}) and (\ref{cloparam}). We will come back to this
point later. For the dual scalars $\varphi_m$ one finds
correspondingly
 \bea
  \delta_{\xi}\varphi_m = \xi^{\rho}\partial_{\rho}\varphi_m
  +2 g\varphi_{(m}\partial_{n)}\xi^{\rho}A_{\rho}^n\;.
 \eea

Let us now check if the covariant derivatives (\ref{covder})
transform covariantly also under the deformed spin-2
transformations. For this it will prove to be convenient to
consider a particular combination of a local $\hat{v}_d$
transformation and a spin-2 transformation. We consider a
diffeomorphism generated by $\xi^{\mu}$ and add a local
$\hat{v}_d$ transformation with field-dependent parameter
$\xi^m=-\xi^{\rho}A_{\rho}^m$. This results in
 \bea
  \begin{split}
   \delta_{\xi}\phi_m^{\alpha}&=\xi^{\rho}D_{\rho}\phi_m^{\alpha}\;,
   \qquad \delta_{\xi}\varphi_m=\xi^{\rho}D_{\rho}\varphi_m\;,
   \\
   \delta_{\xi}e_{\mu}^a &=\xi^{\rho}D_{\rho}e_{\mu}^a+D_{\mu}\xi^{\rho}e_{\rho}^a
   \;,\quad
   \delta_{\xi}A_{\mu}^m = \xi^{\rho}F_{\rho\mu}^m\;,
  \end{split}
 \eea
where we have introduced a covariant derivative on the
transformation parameter $\xi^{\rho}$,
 \bea
  D_{\mu}\xi^{\rho} =
  \partial_{\mu}\xi^{\rho}-gA_{\mu}^m\partial_m\xi^{\rho}\;.
 \eea
We see that except for the gauge field $A_{\mu}^{m}$ the fields
transform like covariant tensors under a gauged notion of
diffeomorphisms, where all partial derivatives have been replaced
by covariant derivatives with respect to $\hat{v}_d$. Moreover, an
action which is constructed out of a density that transforms
accordingly under gauged diffeomorphisms,
 \bea
   \delta_{\xi}(e{\cal L}) = D_{\rho}(e\xi^{\rho}{\cal L})\;,
 \eea
is invariant if and only if it is also invariant under local
$\hat{v}_d$ transformations. This can be shown in complete analogy
to \cite{Hohm:2005sc}. Unfortunately, in contrast to the `bare'
fields, the covariant derivatives of the latter do not transform
covariantly under these gauged diffeomorphisms. This is due to the
non-commutativity of $\hat{v}_d$ covariant derivatives. In fact,
it can be easily checked that
 \bea
  \begin{split}
   [D_{\mu},D_{\nu}]\phi_m^{\alpha} &= -g\partial_n\phi_m^{\alpha}
   F_{\mu\nu}^n-g\phi_n^{\alpha}\partial_mF_{\mu\nu}^n\;, \\
   [D_{\mu},D_{\nu}]e_{\rho}^a &= -g F_{\mu\nu}^n\partial_n
   e_{\rho}^a - g e_{\rho}^a\partial_mF_{\mu\nu}^m\;.
  \end{split}
 \eea
With these relations it can be shown that, e.g.,
 \bea
  \delta_{\xi}(D_{\mu}\phi_m^{\alpha}) =
  \xi^{\rho}D_{\rho}(D_{\mu}\phi_m^{\alpha})+D_{\mu}\xi^{\rho}
  D_{\rho}\phi_m^{\alpha}-g \phi_n^{\alpha}\partial_m\xi^{\rho}F_{\rho\mu}^n\;,
 \eea
i.e. it appears an additional term proportional to the field
strength. This in turn implies that generic $\hat{v}_d$-covariant
actions will not be invariant under the deformed (or gauged)
diffeomorphisms. On the other hand, Kaluza-Klein theories provide
by construction a resolution of these consistency problems. We
will see that they can indeed be written in a manifestly
$\hat{v}_d$-covariant form, i.e.~with all appearing derivatives
being covariant in the sense defined above. At the same time they
are invariant under the spin-2 transformations defined in
(\ref{diffaction}) and (\ref{complor}), the latter fact just
expressing the diffeomorphism invariance of the original
Einstein-Hilbert action. However, inspecting (\ref{diffaction})
and (\ref{complor}) more closely, we infer that this can only be
achieved by adding a scalar-dependent term to the variation
$\delta_{\xi}A_{\mu}^m$. Moreover, we will see that the full
action contains additional terms, beyond those resulting just from
a minimal substitution in (\ref{ungaugeddual}). Those couplings
will contain a scalar potential and spin-2 mass terms, and the
total variation of the action under the gauged diffeomorphisms
will therefore link all terms in a non-trivial way. This should be
compared to the gauging of supergravity, where the supersymmetry
variations also have to be extended by terms proportional to the
gauge-coupling. So the invariance under gauged diffeomorphisms,
which is guaranteed by construction, is not manifest, but we are
going to prove in the next section that on the subsector of purely
gravitational and gauge fields this invariance can be even made
manifest via a Chern-Simons description.

\subsection{Chern-Simons theory for gravitational and gauge
fields}\label{CSsec} As we have discussed in the last section,
simply covariantising a spin-2 invariant action with respect to
the internal diffeomorphism algebra $\hat{v}_d$ in general does
not result in an action which is invariant under any
covariantiation of the spin-2 symmetries. This drawback holds also
for the Einstein-Hilbert term, which in turn is the reason that
the spin-2 graviton usually cannot be charged with respect to some
gauge group. Generalizing the results of \cite{Hohm:2005sc} and in
analogy to gauged supergravity we are going to show that in the
$2+1$ dimensional context consistency can be regained by adding a
Chern-Simons term for the gauge fields. The pure Einstein-Hilbert
term in the ungauged phase -- still depending on all $D$
coordinates -- has, on the other hand, also an interpretation as a
Chern-Simons theory, where the gauge group is given by the current
group $T^d ISO(1,2)$ associated to the Poincar\'e group (see
appendix A of \cite{Hohm:2005sc}, which generalizes
\cite{Witten:1988hc}). Thus, following \cite{Hohm:2005sc}, one
might hope to be able to combine these Chern-Simons terms into one
Chern-Simons theory for some extended Lie algebra. In this spirit
the question of a consistent extension of the `gauged'
Einstein-Hilbert term translates into a purely algebraic problem.
Namely, it has to be shown that a consistent Lie algebra exists,
which has the following properties: First it has to be a
semi-direct product between $T^d iso(1,2)$ and $\hat{v}_d$, which
gives rise to the correct transformation properties under
$\hat{v}_d$ given in (\ref{diffaction}). Second, there has to
exist an invariant non-degenerate quadratic form, which can be
used to construct a Chern-Simons action which is gauge-invariant,
and whose equations of motion do not degenerate. It turns out that
these requirements can indeed be satisfied.

First of all we have to note that for the $\hat{v}_d$ generators
$Q^{m[\underline{j}]}$ alone an invariant bilinear form does not
exist. Thus, as in \cite{Hohm:2005sc}, we have to extend the Lie
algebra further by adding generators $e^{m[\underline{j}]}$, which
transform under $\hat{v}_d$ in such a way that the bilinear
expression $Q^{m[\underline{j}]}e^{m[-\underline{j}]}$ is
invariant. These are exactly given by the `shift' generators of
$T^d sl(d+1,\mathbb{R})$ in (\ref{semialg}). Put differently, we
have to gauge not only $\hat{v}_d$, but the entire subalgebra of
the rigid symmetry in sec.~\ref{rigid} which is spanned by
$Q^{m[\underline{j}]}$ and $e^{m[\underline{j}]}$.

The Lie algebra consistent with the above requirements reads
 \bea\label{bigalgebra}
  \begin{split}
   [P_a^{[\underline{j}]},J_b^{[\underline{k}]}] &= \varepsilon_{abc}
   P^{c[\underline{j}+\underline{k}]}+i\alpha k_p\eta_{ab}e^{p[\underline{j}
   +\underline{k}]}\;, \\
   [J_a^{[\underline{j}]},J_b^{[\underline{k}]}] &= \varepsilon_{abc}
   J^{c[\underline{j}+\underline{k}]}\;, \qquad
   [P_a^{[\underline{j}]},P_b^{[\underline{k}]}] = 0\;, \\
   [Q^{m[\underline{j}]},Q^{n[\underline{k}]}] &= ig
   \left(j_nQ^{m[\underline{j}+\underline{k}]}
   -k_mQ^{n[\underline{j}+\underline{k}]}\right) \;, \\
   [Q^{m[\underline{j}]},P_a^{[\underline{k}]}]&= ig(-j_m-k_m)
   P_a^{[\underline{j}+\underline{k}]}\;, \\
   [Q^{m[\underline{j}]},J_a^{[\underline{k}]}]&= -igk_m
   J_a^{[\underline{j}+\underline{k}]}\;, \\
   [Q^{m[\underline{j}]},e^{n[\underline{k}]}]&=-ig\left( (j_m+k_m)
   e^{n[\underline{j}+\underline{k}]}+\delta^{mn}j_l
   (e^l)^{[\underline{j}+\underline{k}]}\right)\;, \\
   [P_a^{m[\underline{j}]},e^{n[\underline{k}]}]
   &=[J_a^{m[\underline{j}]},e^{n[\underline{k}]}]=
   [e^{m[\underline{j}]},e^{n[\underline{k}]}]=0\;,
  \end{split}
 \eea
and leaves only one free parameter $\alpha$. Here $P_a$ and $J_a$
denote the Poincar\'e generator. We observe like in
\cite{Hohm:2005sc} that the $e^{m[\underline{j}]}$ act as central
extensions for the Poincar\'e subalgebra (or as non-central
extensions for the full algebra). As required, this algebra
carries an invariant non-degenerate bilinear form. In the basis
(\ref{bigalgebra}) it reads
 \bea\label{quadform}
  \langle P_a^{[\underline{j}]},J_b^{[\underline{k}]}\rangle =
  \eta_{ab}\delta^{[\underline{j}],[\underline{k}]}\;, \qquad
  \langle Q^{m[\underline{j}]}, e^{n[\underline{k}]}\rangle =
  \frac{g}{\alpha}\delta^{mn}\delta^{[\underline{j}],[\underline{k}]}\;,
 \eea
while all other terms vanish. Here we have introduced the
short-hand notation
$\delta^{[\underline{j}],[\underline{k}]}=\delta^{j_1,k_1}...\hspace{0.2em}\delta^{j_d,k_d}$.
Note that the invariance of this quadratic form is only insured
due to the central extension of the Poincar\'e subalgebra.

As a next step we are going to construct the Chern-Simons action
associated to the Lie algebra (\ref{bigalgebra}). The action for a
Lie algebra valued gauge field ${\cal A}_{\mu}$ is given by
 \bea\label{CSaction}
  \begin{split}
   S_{\rm CS} &=  \int {\rm Tr}\big( {\cal A}\wedge d{\cal
   A}+\frac{2}{3}{\cal A}\wedge{\cal A}\wedge{\cal A}\big) \\
   &=\frac{1}{2}\int d^3 x \hspace{0.2em}\varepsilon^{\mu\nu\rho}\big(\langle
   {\cal A}_{\mu},\partial_{\nu}{\cal A}_{\rho}-\partial_{\rho}
   {\cal A}_{\nu}\rangle
   +\frac{2}{3}\langle {\cal A}_{\mu},[{\cal A}_{\nu},
   {\cal A}_{\rho}]\rangle \big)  \;,
  \end{split}
 \eea
where the trace is a symbolic notation for an invariant quadratic
form, used explicitly in the second line of (\ref{CSaction}).
Writing the gauge field as
 \bea
  {\cal
  A}_{\mu}=e_{\mu}^{a[\underline{j}]}P_a^{[\underline{j}]}+\omega_{\mu}^{a[\underline{j}]}
  J_a^{[\underline{j}]}+A_{\mu}^{m[\underline{j}]}Q^{m[\underline{j}]}
  +B_{\mu\hspace{0.1em}m}^{[\underline{j}]}e^{m[\underline{j}]}\;,
 \eea
and inserting into (\ref{CSaction}) gives by use of
(\ref{bigalgebra}) and (\ref{quadform}) rise to the following
action
 \bea\label{CSEH}
  S_{\rm CS} =\int d^3x \hspace{0.2em}
  \varepsilon^{\mu\nu\rho}\big(e_{\mu}^{a[-\underline{j}]}(D_{\nu}\omega_{\rho
  a}^{[\underline{j}]}-D_{\rho}\omega_{\nu
  a}^{[\underline{j}]}+\varepsilon_{abc}\omega_{\nu}^{b[\underline{j}-\underline{k}]}
  \omega_{\rho }^{c[\underline{k}]})\big) +
  \frac{g}{\alpha}\varepsilon^{\mu\nu\rho}B_{\mu\hspace{0.1em}m}^{[-\underline{j}]}F_{\nu\rho}^{m[\underline{j}]}\;.
 \eea
Thus, we exactly recover the covariantized Einstein-Hilbert term,
in which the partial derivatives have been replaced by covariant
ones with respect to $\hat{v}_d$. In addition we get a
Chern-Simons term for the gauge fields.

Let us now discuss the equations of motion and the symmetries of
this Chern-Simons theory. Varying (\ref{CSaction}) with respect to
the gauge field ${\cal A}_{\mu}$ yields vanishing field strength
as the equations of motion. For (\ref{CSEH}) this implies
 \bea\label{eom0}
  {\cal F}_{\mu\nu}=R_{\mu\nu}^{a[\underline{j}]}J_a^{[\underline{j}]}
  +T_{\mu\nu}^{a[\underline{j}]}P_a^{[\underline{j}]}
  +F_{\mu\nu}^{m[\underline{j}]}Q^{m[\underline{j}]}+G_{\mu\nu\hspace{0.2em}m}
  ^{[\underline{j}]}e^{m[\underline{j}]} = 0\;,
 \eea
whose components read
 \begin{eqnarray}\label{eom}
   R_{\mu\nu}^{a[\underline{j}]} &=&
   \partial_{\mu}\omega_{\nu}^{a[\underline{j}]}-
   \partial_{\nu}\omega_{\mu}^{a[\underline{j}]}+\varepsilon^{abc}
   \omega_{\mu b}^{[\underline{j}-\underline{k}]}\omega_{\nu c}
   ^{[\underline{k}]} \\ \nonumber
   &&+ ig(j_n-k_n)\omega_{\mu}^{a[\underline{j}-\underline{k}]}A_{\nu}
   ^{n[\underline{k}]}-igk_mA_{\mu}^{m[\underline{j}-\underline{k}]}
   \omega_{\nu}^{a[\underline{k}]}\;, \\ \nonumber
   T_{\mu\nu}^{a[\underline{j}]} &=& \partial_{\mu}e_{\nu}^{a[\underline{j}]}
   -\partial_{\nu}e_{\mu}^{a[\underline{j}]}+\varepsilon^{abc}e_{\mu b}
   ^{[\underline{j}-\underline{k}]}\omega_{\nu c}^{[\underline{k}]}
   +\varepsilon^{abc}\omega_{\mu b}
   ^{[\underline{j}-\underline{k}]}e_{\nu c}^{[\underline{k}]} \\ \nonumber
   &&+igj_ne_{\mu}^{a[\underline{j}-\underline{k}]}A_{\nu}^{n[\underline{k}]}
   -igj_m A_{\mu}^{m[\underline{j}-\underline{k}]}e_{\nu}^{a[\underline{k}]}\;,
   \\ \nonumber
   F_{\mu\nu}^{m[\underline{j}]} &=& \partial_{\mu}A_{\nu}^{m[\underline{j}]}
   -\partial_{\nu}A_{\mu}^{m[\underline{j}]}+ig(j_n-k_n)
   A_{\mu}^{m[\underline{j}-\underline{k}]}A_{\nu}^{n[\underline{k}]}
   -igk_nA_{\mu}^{n[\underline{j}-\underline{k}]}A_{\nu}^{m[\underline{k}]}\;,
   \\ \nonumber
   G_{\mu\nu\hspace{0.1em}m}^{[\underline{j}]} &=& \partial_{\mu}
   B_{\nu\hspace{0.1em}m}-\partial_{\nu}B_{\mu\hspace{0.1em}m}
   +i\alpha k_me_{\mu}^{a[\underline{j}-\underline{k}]}\omega_{\nu a}
   ^{[\underline{k}]}-i\alpha(j_m-k_m)\omega_{\mu}
   ^{a[\underline{j}-\underline{k}]}e_{\nu a}^{[\underline{k}]} \\ \nonumber
   &&+ igj_n B_{\mu\hspace{0.1em}m}^{[\underline{j}-\underline{k}]}
   A_{\nu}^{n[\underline{k}]}+igk_mB_{\mu\hspace{0.1em}n}
   ^{[\underline{j}-\underline{k}]}A_{\nu}^{n[\underline{k}]} \\ \nonumber
   &&- igj_nA_{\mu}^{n[\underline{j}-\underline{k}]}B_{\nu\hspace{0.1em}m}
   ^{[\underline{k}]}-ig(j_m-k_m)A_{\mu}^{n[\underline{j}-\underline{k}]}
   B_{\nu\hspace{0.1em}n}^{[\underline{k}]}\;.
 \end{eqnarray}
Like the covariant derivatives also the field strength given here
can be conveniently rewritten by taking the fields to be dependent
on all $D$ coordinates and writing the non-abelian contributions
in (\ref{eom}) by means of an internal derivative $\partial_m$.
This results in
 \begin{eqnarray}\label{curv}
  R_{\mu\nu}^a &=& D_{\mu}\omega_{\nu}^a-D_{\nu}\omega_{\mu}^a
  +\varepsilon^{abc}\omega_{\mu b}\omega_{\nu c}\;,
  \\ \nonumber
  T_{\mu\nu}^a &=& D_{\mu}e_{\nu}^a - D_{\nu}e_{\mu}^a + \varepsilon^{abc}e_{\mu b}\omega_{\nu c}
  +\varepsilon^{abc}\omega_{\mu b}e_{\nu c}\;,  \\ \nonumber
  F_{\mu\nu}^m&=&\partial_{\mu}A_{\nu}^m-\partial_{\nu}A_{\mu}^m
  -gA_{\mu}^n\partial_nA_{\nu}^m+gA_{\nu}^n\partial_nA_{\mu}^m\;,
  \\ \nonumber
  G_{\mu\nu\hspace{0.1em}m}&=&\partial_{\mu}B_{\nu\hspace{0.1em} m}-\partial_{\nu}B_{\mu\hspace{0.1em}m}
  -2g A_{[\mu}^n\partial_n B_{\nu]m}+2gB_{[\mu
  n}\partial_mA_{\nu]}^n\\ \nonumber
  &&+2gB_{[\mu m}\partial_nA_{\nu ]}^n+2\alpha e_{[\mu}^a\partial_m
  \omega_{\nu ]a}\;. \nonumber
 \end{eqnarray}
We observe in particular that the standard formulas for the
Riemann tensor $R_{\mu\nu}^a$ (in three dimensions) and for the
torsion tensor $T_{\mu\nu}^a$ are recovered, but with all
derivatives being $\hat{v}_d$ covariant. Let us note that the
torsion constraint $T_{\mu\nu}^a =0$ following from the equations
of motion (\ref{eom0}) can be used as in standard gravity to solve
for the spin connection $\omega_{\mu}^a$, but here in terms of
both $e_{\mu}^a$ and $A_{\mu}^m$.

In order to analyze the symmetries of (\ref{CSEH}) and thus of
(\ref{eom}), let us consider the non-abelian gauge transformations
determined by (\ref{bigalgebra}). Even though the Chern-Simons
action is not manifestly gauge invariant, it can be easily checked
that up to a total derivative it is invariant under the gauge
transformations $\delta{\cal
A}_{\mu}=D_{\mu}u=\partial_{\mu}u+[{\cal A}_{\mu},u]$ generated by
a Lie algebra valued transformation parameter $u$. Writing the
transformation parameter as
 \bea
  u=\rho^{a[\underline{j}]}P_a^{[\underline{j}]}+\tau^{a[\underline{j}]}
  J_a^{[\underline{j}]}+\xi^{n[\underline{j}]}Q^{n[\underline{j}]}
  +\Lambda_n^{[\underline{j}]}e^{n[\underline{j}]}\;,
 \eea
the gauge transformations are given by
 \begin{eqnarray}\label{YMvar}
  \delta e_{\mu}^{a[\underline{j}]} &=& \partial_{\mu}\rho^{a[\underline{j}]}
  +\varepsilon^{abc}e_{\mu b}^{[\underline{j}-\underline{k}]}
  \tau_c^{[\underline{j}]}+\varepsilon^{abc}
  \omega_{\mu b}^{[\underline{j}-\underline{k}]}\rho_c^{[\underline{k}]}
   \\ \nonumber
  &&+ igj_n\xi^{n[\underline{k}]}e_{\mu}^{a[\underline{j}-\underline{k}]}
  -igj_m\rho^{a[\underline{k}]}A_{\mu}^{m[\underline{j}-\underline{k}]}
  \;, \\ \nonumber
  \delta \omega_{\mu}^{a[\underline{j}]} &=& \partial_{\mu}
  \tau^{a[\underline{j}]}+\varepsilon^{abc}
  \omega_{\mu b}^{[\underline{j}-\underline{k}]}
  \tau_c^{[\underline{k}]} 
  + ig(j_n-k_n)\xi^{n[\underline{k}]}
  \omega_{\mu}^{a[\underline{j}-\underline{k}]}
  -igk_m\tau^{a[\underline{k}]}A_{\mu}^{m[\underline{j}-\underline{k}]}\;,
  \\ \nonumber
  \delta A_{\mu}^{m[\underline{j}]} &=&
  \partial_{\mu}\xi^{m[\underline{j}]}+ig(j_n-k_n)\xi^{n[\underline{k}]}A_{\mu}^{m[\underline{j}-\underline{k}]}
  -igk_n\xi^{m[\underline{k}]}A_{\mu}^{n[\underline{j}-\underline{k}]}\;,
  \\ \nonumber
  \delta B_{\mu\hspace{0.1em}m}^{[\underline{j}]} &=& \partial_{\mu}\Lambda_m^{[\underline{j}]}
  +i\alpha k_m
  e_{\mu}^{a[\underline{j}-\underline{k}]}\tau_a^{[\underline{k}]}-i\alpha
  (j_m-k_m)\omega_{\mu}^{a[\underline{j}-\underline{k}]}\rho_a^{[\underline{k}]}
  -igj_nA_{\mu}^{n[\underline{j}-\underline{k}]}\Lambda_m^{[\underline{k}]}
  \\ \nonumber
  &&- ig(j_m-k_m)A_{\mu}^{n[\underline{j}-\underline{k}]}
  \Lambda_n^{[\underline{k}]}+igj_n
  B_{\mu\hspace{0.1em}m}^{[\underline{j}-\underline{k}]}\xi^{n[\underline{k}]}
  +igk_mB_{\mu\hspace{0.1em}n}^{[\underline{j}-\underline{k}]}\xi^{n[\underline{k}]}\;.
 \end{eqnarray}
Also the gauge transformations can be conveniently rewritten by
taking $y$-dependent fields and transformation parameters. The
result reads
 \begin{eqnarray}\label{yvar}
   \delta e_{\mu}^a & = &
   \partial_{\mu}\rho^a+\epsilon^{abc}e_{\mu
   b}\tau_c+\epsilon^{abc}\omega_{\mu
   b}\rho_c+g\xi^n\partial_ne_{\mu}^a+g\partial_n\xi^n e_{\mu}^a \\
   \nonumber
   &&-g\rho^a\partial_m A_{\mu}^m-g\partial_m\rho^a A_{\mu}^m\;, \\
   \nonumber
   \delta\omega_{\mu}^a &=&
   \partial_{\mu}\tau^a+\epsilon^{abc}\omega_{\mu
   b}\tau_c+g\xi^n\partial_n\omega_{\mu}^a-gA_{\mu}^m\partial_m\tau^a\;,
   \\
   \nonumber
   \delta A_{\mu}^m &=&
   \partial_{\mu}\xi^m+g\xi^n\partial_nA_{\mu}^m-gA_{\mu}^n\partial_n\xi^m\;,
   \\
   \nonumber
   \delta B_{\mu\hspace{0.1em}m}& = & \partial_{\mu}\Lambda_m -
   g\Lambda_m\partial_n A_{\mu}^n -
   gA_{\mu}^n\partial_n\Lambda_m-g\Lambda_n\partial_mA_{\mu}^n \\
   \nonumber
   &&+g\xi^n\partial_n B_{\mu\hspace{0.1em}m}+gB_{\mu\hspace{0.1em}m}\partial_n\xi^n +
   g\partial_m\xi^n B_{\mu\hspace{0.1em}n} + \alpha
   e_{\mu}^a\partial_m\tau_a-\alpha
   \rho_a\partial_m\omega_{\mu}^a\;.
 \end{eqnarray}
Let us now check, if the symmetries expected for Kaluza-Klein
theories from (\ref{diffaction}) and (\ref{complor}) are contained
in (\ref{yvar}). First of all we notice by comparing with
(\ref{diffaction}), that the $\hat{v}_d$ gauge transformations
parametrized by $\xi^m$ are correctly reproduced. Moreover,
(\ref{yvar}) allows us to compare with the deformed spin-2
transformations, or in other words, to see if one recovers the
gauged diffeomorphisms. For this we consider the non-abelian gauge
transformations for the field-dependent transformation parameter
 \bea
  \rho^a = \xi^{\mu}e_{\mu}^a\;, \qquad \tau^a =
  \xi^{\mu}\omega_{\mu}^a\;, \qquad \xi^m = \xi^{\mu}A_{\mu}^m\;,
  \qquad \Lambda_m = \xi^{\mu}B_{\mu\hspace{0.1em}m}\;.
 \eea
Then the gauge transformations (\ref{yvar}) take the following
form
 \begin{eqnarray}\label{gaugetr}
  \delta e_{\mu}^a &=&
  \xi^{\rho}\partial_{\rho}e_{\mu}^a+\partial_{\mu}\xi^{\rho}e_{\rho}^a
  +gA_{\rho}^m\partial_m\xi^{\rho}e_{\mu}^a-gA_{\mu}^m\partial_m\xi^{\rho}e_{\rho}^a
  -\xi^{\rho}T_{\rho\mu}^a\;, \\ \nonumber
  \delta A_{\mu}^m &=& \xi^{\rho}\partial_{\rho}A_{\mu}^m +
  \partial_{\mu}\xi^{\rho}e_{\rho}^a-gA_{\mu}^n\partial_n\xi^{\rho}A_{\rho}^m
  -\xi^{\rho}F_{\rho\mu}^m\;.  \nonumber
 \end{eqnarray}
We see that up to field strength terms, which vanish by the
equations of motion (\ref{eom0}), the deformed spin-2
transformations are correctly reproduced for $e_{\mu}^a$, and for
$A_{\mu}^m$ up to scalar-dependent terms. Thus on-shell the spin-2
variations are contained in the non-abelian gauge transformations.
Even though (\ref{yvar}) is only on-shell equivalent to
(\ref{diffaction}) and (\ref{complor}), the gauged diffeomorphisms
are also an (off-shell) symmetry. This can be checked explicitly,
but follows also from the fact that gauge transformations in
general can be `twisted' by terms proportional to the equations of
motion \cite{Julia:1999tk}. In fact, a symmetry $\delta\phi^i$
(where $\phi^i$ generically denotes the fields) can be rewritten
by use of a so-called trivial gauge transformation,
 \bea\label{trivial}
  \bar{\delta}\phi^i = \delta\phi^i + \Omega^{ij}E_j\;,
 \eea
where $E_i = \delta{\cal L}/\delta\phi^i$, if the (space-time
dependent) $\Omega^{ij}$ are anti-symmetric. If we choose in case
of the Chern-Simons theory this matrix to be
$\Omega_{\mu\nu}=\frac{1}{2}\epsilon_{\mu\nu\rho}\xi^{\rho}$ the
twisted gauge transformation reads by use of
$E^{\mu}=\epsilon^{\mu\nu\rho}{\cal F}_{\nu\rho}$
 \bea
  \bar{\delta}{\cal A}_{\mu} = \delta{\cal A}_{\mu} +
  \Omega_{\mu\nu}E^{\nu} = \delta{\cal A}_{\mu} + \xi^{\rho}{\cal
  F}_{\rho\mu}\;,
 \eea
i.e.~it receives the term proportional to the field strength in
(\ref{gaugetr}).

In total we have found an action which is invariant under deformed
spin-2 gauge transformations or, equivalently, under gauged
diffeomorphisms, with the latter being realized as ordinary
Yang-Mills gauge transformations. This has been achieved by virtue
of a Chern-Simons formulation based on a centrally extended
Poincar\'e algebra. In the next section we will turn to the
problem of reconstructing the full Einstein-Hilbert action.

\subsection{The full theory}
Let us now discuss the problem of gauging the internal
diffeomorphism algebra $\hat{v}_d$ for the scalar fields. Also for
this we first have to replace the partial derivatives by the
covariant ones defined in (\ref{covder}). However, in the last
section we have seen that in order to get a consistent
covariantisation of the Einstein-Hilbert term we have to introduce
additional gauge fields $B_{\mu\hspace{0.1em}m}$, which gauge the
shift symmetries of $SL(d+1,\mathbb{R})$. Since these shift
symmetries act also on the scalars $\varphi_m$ according to
$\delta_{\Lambda}\varphi_m= -g\Lambda_m$, this implies that the
covariant derivative for the latter has to be extended to
 \bea
  {\cal D}_{\mu}\varphi_m = D_{\mu}\varphi_m +gB_{\mu
  \hspace{0.1em}m}\;.
 \eea

Performing this minimal substitution in (\ref{ungaugeddual}),
results in an action of the form
 \bea\label{covaction}
  \begin{split}
   S_{g} = \int d^3x d^dy \hspace{0.12em}e\big(&-R_3^{\rm cov} - \frac{1}{2}g
   e^{-1}\varepsilon^{\mu\nu\rho}B_{\mu\hspace{0.1em}m}F_{\nu\rho}^m
   +\phi^{-2}D^{\mu}\phi D_{\mu}\phi \\
   &+\frac{1}{2}\phi^{-2}G^{mn}(\phi)
   {\cal D}^{\mu}\varphi_m{\cal D}_{\mu}\varphi_n
   + \frac{1}{2}(\phi_{\alpha}^mD^{\mu}\phi_m^{\gamma})
   (\phi_{\gamma}^nD_{\mu}\phi_n^{\alpha})\\
   &-\frac{1}{2}G^{mn}(\phi)
   D^{\mu}\phi_m^{\beta}D_{\mu}\phi_{n\beta}
   + \hspace{0.2em}{\cal L}_{\rm gauge} \big)\;.
  \end{split}
 \eea
Here $R^{\rm cov}_3$ denotes the (2+1)-dimensional
Einstein-Hilbert term, covariantized with respect to the local
$\hat{v}_d$ symmetry, as introduced in (\ref{CSEH}). Moreover, we
added the Chern-Simons term for the gauge vectors (setting
$\alpha=2$ in (\ref{CSEH})), which is required according to the
analysis in the previous section. Finally we supplemented the
action by additional couplings ${\cal L}_{\rm gauge}$ which, by
the experience from $S^1$ theories and gauged supergravities
\cite{Nicolai:2000sc,Nicolai:2001sv}, are expected to appear.

In fact, facing the problem whether (\ref{covaction}) is invariant
under the $g$-deformed spin-2 transformations introduced in
\ref{covsection}, we have to conclude that this is in general not
the case, since the consistency problems have been resolved only
for the topological subsector consisting of gravitational and
gauge fields. This in turn is the reason that the transformation
rules and couplings have to be extended further. We do not aim to
determine all possible couplings systematically in this paper, but
instead prove that those resulting from a direct Kaluza-Klein
analysis fit exactly into a theory of the form given in
(\ref{covaction}). More precisely, we show that upon choosing
${\cal L}_{\rm gauge}$ in (\ref{covaction}) as determined by the
Kaluza-Klein approach in appendix B one gets an action, which is
on-shell equivalent to the full Kaluza-Klein theory and whose
invariance under spin-2 transformations can be traced back to the
invariance of the original Kaluza-Klein theory.

First we have to show that (\ref{covaction}) with the Chern-Simons
gauge fields is equivalent to the Kaluza-Klein theory containing
Yang-Mills terms. This can be seen in the same way as in gauged
supergravity, following \cite{Nicolai:2003bp,deWit:2003ja}. In
fact, varying (\ref{covaction}) with respect to
$B_{\mu\hspace{0.1em}m}$ results in
 \bea
  {\cal D}_{\mu}\varphi_m =
  \frac{1}{2}\phi^2G_{mn}\varepsilon_{\mu\nu\rho}F^{\nu\rho\hspace{0.1em}n}\;,
 \eea
which in the ungauged limit $g\rightarrow 0$ reduces to the
standard duality relation (\ref{dualrel}). This in turn implies
that the equations of motion for (\ref{covaction}) are equivalent
to those in the Yang-Mills gauged form (\ref{EHYMfull}), which can
be most easily seen by choosing the gauge fixing $\varphi_m =0$
and then integrating out $B_{\mu\hspace{0.1em}m}$. Finally, we
know from (\ref{curv}) that varying (\ref{covaction}) with respect
to $\omega_{\mu}^a$ results in the $\hat{v}_d$ covariantized
torsion constraint. Using the latter to solve for $\omega_{\mu}^a$
in terms of $e_{\mu}^a$ and $A_{\mu}^m$, one gets an $\hat{v}_d$
covariantized Einstein-Hilbert term, which coincides exactly with
that appearing upon direct dimensional reduction. Indeed, in
appendix B we prove that the Kaluza-Klein theory has the required
form, where ${\cal L}_{\rm gauge}$ is given by
 \bea\label{gaugeind}
  \begin{split}
   {\cal L}_{\rm gauge} = &-\frac{1}{2}g^2\phi^{-2}G^{mn}(\phi)(e_a^{\nu}D_me_{\nu}^b)
   (e_b^{\mu}D_ne_{\mu}^a)-\frac{1}{2}g^2\phi^{-2}G^{mn}(\phi)g^{\mu\nu} D_me_{\mu}^aD_n e_{\nu
   a}\\
   &-g^2\phi^{-2}G^{mn}(e^{a\mu}D_m e_{\mu a})(e^{b\nu}D_n e_{\nu b})
   +g^2\phi^{-2}R(\phi)-\frac{1}{2}gF^{ab\hspace{0.2em}m}e_{[a}^{\nu}D_me_{\nu
   b]}\;.
  \end{split}
 \eea
Here $R(\phi)$ denotes the Ricci scalar computed with respect to
the internal vielbein $\phi_m^{\alpha}$ in the standard fashion,
and $D_me_{\mu}^a$ is defined in (\ref{Dm}). The terms quadratic
in $D_me_{\mu}^a$ are mass terms for the spin-2 fields, while
$R(\phi)$ is a scalar potential. With (\ref{gaugeind}) we note
that in the decompactification limit $g\rightarrow 0$ the theory
indeed reduces to (\ref{ungaugedaction}).

After we have proven that (\ref{covaction}) is equivalent to the
$D$-dimensional Einstein-Hilbert action, we can conclude that it
admits, at least on-shell, the gauged diffeomorphisms discussed in
\ref{covsection} (after adding the scalar-dependent contribution
in $\delta_{\xi}A_{\mu}^m$) as a local symmetry. However, in
contrast to the pure Chern-Simons theory in \ref{CSsec}, the
symmetry on the remaining couplings is far from being manifest.
The difficulty in analyzing this symmetry is due to the fact that
the original theory is in a 2$^{\rm nd}$ order form, in which the
spin connection $\omega_{\mu}^a$ and the dual vector
$B_{\mu\hspace{0.1em}m}$ do not appear as independent fields, but
are determined by their equations of motion, while the scalars
$\varphi_m$ are altogether absent. In the following let us
therefore briefly comment on the different realization of
symmetries in a 1$^{\rm st}$ order and a 2$^{\rm nd}$ order
formulation.

Concerning the problem to find an independent symmetry variation
for the spin connection, we note that $\omega_{\mu}^a$, when
expressed in terms of $e_{\mu}^a$ and $A_{\mu}^m$, does not
transform simply as in the 1$^{\rm st}$ order Chern-Simons
formulation in (\ref{yvar}), but in a highly non-trivial manner.
The latter fact can be traced back to the same origin as the
non-covariance of $\hat{v}_d$-covariant derivatives under gauged
diffeomorphisms discussed in sec.~\ref{covsection}. It is
nevertheless always possible for a given 2$^{\rm nd}$ order action
with a certain local symmetry and an on-shell equivalent 1$^{\rm
st}$ order action to find a corresponding local (off-shell)
symmetry on the 1$^{\rm st}$ order fields (for a systematic
account see \cite{Julia:1999tk}). But, it has to be taken into
account that these variations in general cannot just be determined
by applying the 2$^{\rm nd}$ order variation to, in our case,
$\omega_{\mu}^a(e,A)$, but they receive additional contributions
which, however, vanish on-shell (see, e.g., eq.~(2.7) in
ref.~\cite{Julia:1999tk}). Even more, the 1$^{\rm st}$ order
formulation is not unique since trivial gauge transformations as
in (\ref{trivial}) can be added, which in turn can simplify the
expressions significantly. One may compare with the situation in
supergravity. Applying naively the mentioned results of
\cite{Julia:1999tk} in order to get a 1$^{\rm st}$ order
formulation of pure $N=1$ supergravity in $D=4$
\cite{Deser:1976eh} results in a rather intricate supersymmetry
variation for the spin connection. Only after a trivial gauge
transformation they take a much simpler, namely supercovariant
form, in the sense that the variations do not contain derivatives
of the supersymmetry parameter. Similarly, one may hope to find a
true 1$^{\rm st}$ order formulation of (\ref{covaction}), which
treats both $\omega_{\mu}^a$ and $B_{\mu\hspace{0.1em}m}$ as
independent fields, and which takes advantage of the Chern-Simons
formulation of sec.~\ref{CSsec}. We will leave this for future
work.

\section{Conclusions and Outlook}
In this paper we further analyzed the local spin-2 and global
hidden symmetries that appear in Kaluza-Klein theories once all
massive modes are taken into account, generalizing
\cite{Hohm:2005sc} to the case of generic internal manifolds. We
found that in the unbroken phase the hidden rigid symmetry group
$SL(D-2,\mathbb{R})$, that appears by reducing $D$-dimensional
gravity on a torus to three dimensions, is enhanced to the
infinite-dimensional current group associated to the internal
manifold $K_d$. Moreover, the diffeomorphism algebra of the
internal manifold, generalizing the Virasoro algebra in the case
of a circle reduction, shows up as a global symmetry. The broken
phase in turn results from a gauging of $\hat{v}_d$ and a certain
subalgebra of $T^d sl(d+1,\mathbb{R})$. We proved that the spin-2
and spin-1 fields can be incorporated into an action of
Chern-Simons form and we gave the underlying gauge algebra
explicitly in case of a torus.

Even though our analysis was restricted to a torus as far as the
mode expansion or gauge algebra is concerned, we can immediately
conclude that the presented picture holds more generally. In fact,
since the action can be entirely rewritten in terms of fields
depending on the internal coordinates $y^m$, without any reference
to the topology of a torus, the unbroken phase
(\ref{ungaugeddual}) as well as the Chern-Simons action
(\ref{CSEH}) exist for any internal manifold $K_d$ and have the
required symmetries. For instance, since the Chern-Simons action
is gauge invariant for any $K_d$, we know that the analogue of the
gauge algebra (\ref{bigalgebra}) has to exist and can be given
explicitly, e.g., simply by expanding the gauge transformations
(\ref{yvar}) in harmonics of $K_d$ and reading off the Lie algebra
from the homogenous terms. Similarly, for the global symmetry
algebra in the unbroken phase the semi-direct product between
$diff(K_d)$ and $K_d sl(d+1,\mathbb{R})$ is completely determined
by (\ref{semi}) and (\ref{transparam}). In fact, once all
Kaluza-Klein modes are taken into account, the resulting theory is
independent of the internal manifold in the sense that the latter
just serves as a reference geometry, which determines the
formulation in terms of harmonics, but does not affect the
physical content (which is, in the present case, still that of
$D$-dimensional Einstein gravity). A formulation in terms of
harmonics is of crucial importance only if truncations are
considered, and one may hope that the knowledge of the
corresponding gauge algebra allows a systematic analysis of
consistent truncations as an algebraic problem (compare
\cite{deWit:1986iy}).

Let us stress again that we have provided a classically equivalent
reformulation of pure gravity in any dimension $D$, since no
truncations were involved. However, the duality transformations
specific for $D=3$ were still possible, and so the physical
degrees of freedom in (\ref{covaction}) are described by the
scalars of a gauged non-linear $\sigma$-model, while the `kinetic
terms' of the graviton modes are given by a topological
Chern-Simons action. This reformulation may therefore enlighten
the meaning of hidden symmetries in terms of the original,
higher-dimensional theory in that the latter can be viewed as a
deformation (in the sense of a gauging) of a theory which has
these symmetries.

This work can be extended into various directions. First of all, a
true 1$^{\rm st}$ order formulation as discussed in the previous
section would clearly be desirable in order to analyze more
systematically which kind of couplings are allowed by gauged
diffeomorphisms. At this stage we know about only one consistent
gauging (namely (\ref{covaction})), since this one has a
higher-dimensional origin. However, it could well be that gaugings
exist, which cannot be derived from known higher-dimensional
theories by the Kaluza-Klein procedure. For instance, this happens
already for ordinary gauged supergravities in $D=3$
\cite{Nicolai:2001sv}. In this case one could not perform the
steps of appendix B to arrive after a dualisation at
(\ref{covaction}), but can only rely on symmetry arguments. Such a
theory would presumably be characterized by more (or different)
gauge symmetries incorporated into the Chern-Simons theory and
another form of the scalar potential. An analysis of this kind
would also be important for the construction of theories which are
not expected to have a formulation as a Lorentz-invariant
gravitational theory, like the 12-dimensional theory proposed in
\cite{Abou-Zeid:1999fv,deWit:2000zz,deWit:2000wu}. Finally, the
$AdS$ and supersymmetric extension would be interesting
\cite{Dolan:1984fm}, e.g.~applied to 11-dimensional supergravity.

\subsection*{Acknowledgments}
I would like to thank Bernard de Wit and Henning Samtleben for
valuable discussions and useful comments.

This work has been supported by the stichting FOM and the European
Union RTN network MRTN-CT-2004-005104.

\begin{appendix}
\renewcommand{\theequation}{\Alph{section}.\arabic{equation}}

\section{Appendix: Non-linear realisation of $SL(d+1,\mathbb{R})$}
\setcounter{equation}{0}\label{A} It is well known that
Kaluza-Klein reduction of (super-)gravities leads to `hidden'
symmetries \cite{Cremmer:1997ct,Cremmer:1999du,Lambert:2001gk}.
The simplest example is the so-called Ehlers group, which appears
upon reducing four-dimensional gravity to $D=3$, or equivalently,
by considering Einstein gravity with one space-like isometry. More
specifically, after dualization into a scalar, the Kaluza-Klein
vector spans together with the dilaton the $\sigma$-model coset
space $SL(2,\mathbb{R})/SO(2)$. This yields the isometry group
$SL(2,\mathbb{R})$ as rigid invariance group, which in turn does
not have an obvious higher-dimensional ancestor. This phenomenon
generalizes to the case of arbitrary torus reductions: Reducing
Einstein gravity on $T^d$ to $D=3$ yields upon dualization a
$\sigma$-model with target space $SL(d+1,\mathbb{R})/SO(d+1)$,
which can be seen as follows.

In order to dualize the Kaluza-Klein vectors $A_{\mu}^m$ into
scalars $\varphi_m$ we enforce as usual the Bianchi identity by
means of Lagrange multipliers, i.e.~the Yang-Mills term in
(\ref{KKaction}) gets replaced by the new action
 \bea
  {\cal L}^{\prime}(F,\varphi)=-\frac{1}{4}\phi^2G_{mn}(\phi)
  F^{\mu\nu\hspace{0.1em}m}F_{\mu\nu}^n-\frac{1}{2}\varphi_m
  \varepsilon^{\mu\nu\rho}\partial_{\mu}F_{\nu\rho}^m\;.
 \eea
Varying with respect to $F_{\mu\nu}^m$ yields the duality
relation
 \bea\label{dualrel}
  \partial_{\mu}\varphi_m = \frac{1}{2}\phi^2 G_{mn}(\phi)
  \varepsilon_{\mu\nu\rho}F^{\nu\rho\hspace{0.1em}m}\;.
 \eea
Integrating out $F_{\mu\nu}^m$ and combining with the
dimensionally reduced action in (\ref{KKaction}) implies
 \bea\label{dualaction}
  \begin{split}
   S=\int d^3x\hspace{0.2em}e\big[-R^3(e)&+g^{\mu\nu}(\phi^{-2}
   \partial_{\mu}\phi\partial_{\nu}\phi+\frac{1}{2}\phi^{-2}G^{mn}(\phi)
   \partial_{\mu}\varphi_m\partial_{\nu}\varphi_n \\
   &+\frac{1}{2}\phi_{\alpha}^m\partial_{\mu}\phi_m^{\gamma}\phi_{\gamma}^n
   \partial_{\nu}\phi_n^{\alpha}-\frac{1}{2}G^{mn}(\phi)\partial_{\mu}
   \phi_m^{\beta}\partial_{\nu}\phi_{n\beta})\big]\;.
  \end{split}
 \eea

To see that this action carries indeed a coset space structure,
let us briefly recall the Lie algebra of $SL(d+1,\mathbb{R})$.
It is convenient to start from the subalgebra $sl(d,\mathbb{R})$.
The latter is a $(d^2-1)$-dimensional algebra which is spanned by
the generators $K^a_{\hspace{0.3em}b}$, $a,b=1,..,d$, satisfying
$K^a_{\hspace{0.3em}a}=0$. The Lie algebra reads
 \bea \label{sld}
  [K_{\hspace{0.3em}b}^a,K_{\hspace{0.3em}d}^c]=
  \delta^a_d K_{\hspace{0.3em}b}^c-\delta^c_b K_{\hspace{0.3em}d}^a \;.
 \eea
An explicit representation by traceless matrices is given by
 \bea\label{sldmatrix}
  (K_{\hspace{0.3em}b}^a)^m_{\hspace{0.3em}n}=\delta_n^a\delta_b^m
  -\frac{1}{d}\delta^a_b\delta^m_n\;.
 \eea
The elements of $sl(d+1,\mathbb{R})$ can then be written as
 \bea
  \hat{K}_{\hspace{0.3em}b}^a=\left(\begin{array}{cc} K_{\hspace{0.3em}b}^a &
  0 \\ 0 & 0 \end{array}\right), \quad
  \hat{e}_a = \left(\begin{array}{cc} 0 & e_a \\ 0 & 0 \end{array}\right),
  \quad
  \hat{f}^a=\left(\begin{array}{cc} 0 & 0 \\ f^a & 0 \end{array}\right),
  \quad
  \hat{e}=\left(\begin{array}{cc} -{\bf 1}_d & 0 \\ 0 & d \end{array}\right),
 \eea
where the components of the column and row vectors are defined by
 \bea
  (e_a)^m = \delta_a^m\;, \qquad (f^a)_m = \delta_m^a \;.
 \eea
The Lie algebra of $sl(d+1,\mathbb{R})$, extending the
$sl(d,\mathbb{R})$ subalgebra (\ref{sld}), can then be easily
computed and turns out to be (after dropping the hats)
 \bea\label{liealgebra}
 \begin{split}
  [K_{\hspace{0.3em}b}^a,e_c] &= \delta_c^a e_b - \frac{1}{d}\delta_b^a e_c\;,
  \qquad
  [K_{\hspace{0.3em}b}^a,f^c] = -\delta^c_b f^a
  + \frac{1}{d}\delta^a_b f^c\;,\\
  [e_a,f^b] &= K_{\hspace{0.3em}a}^b- \frac{1}{d}\hat{e}\delta_a^b\;, \quad
  [K_{\hspace{0.3em}b}^a,\hat{e}] = 0\;. \\
  [e_a,\hat{e}] &= (d+1)e_a \;, \qquad [f^a,\hat{e}] = -(d+1)f^a\;, \\
  [e_a,e_b]&=0\;, \qquad [f_a,f_b]=0\;.
 \end{split}
 \eea

Next we can turn to the construction of the non-linear
$\sigma$-model with target space $SL(d+1,\mathbb{R})/SO(d+1)$. The
scalar fields will be described by a group-valued matrix
 \bea\label{groupelement}
  {\cal V}= \left(\begin{array}{cc} \phi_m^{\alpha} & -\phi^{-1}\varphi_m \\
  0 & \phi^{-1} \end{array}\right)\;,
 \eea
where we have fixed some of the $SO(d+1)$ symmetry to choose a
triangular gauge. Then one can compute the Lie-algebra-valued current
 \bea
  {\cal V}^{-1}\partial_{\mu}{\cal V} = \left(\begin{array}{cc}
  \phi_{\alpha}^m\partial_{\mu}\phi_m^{\beta} & -\phi^{-1}\phi_{\alpha}^m
  \partial_{\mu}\varphi_m \\ 0 & -\phi^{-1}\partial_{\mu}\phi
  \end{array}\right)\;.
 \eea
The $\sigma$-model action can now be defined by decomposing this
current into compact and non-compact parts, i.e.~by decomposing it
into anti-symmetric and symmetric matrices. Denoting the
non-compact part by brackets $[\hspace{0.3em}]$, the resulting
action reads
 \begin{eqnarray}
  {\cal L}_{\rm coset} &=& g^{\mu\nu}{\rm Tr}\left([{\cal V}^{-1}
  \partial_{\mu}{\cal V}][{\cal V}^{-1}\partial_{\nu}{\cal V}]\right)
  \\ \nonumber
  &=& \phi^{-2}\partial^{\mu}\phi\partial_{\mu}\phi +
  \frac{1}{2}\phi^{-2}G^{mn}(\phi)\partial^{\mu}\varphi_m\partial_{\mu}
  \varphi_n + \frac{1}{2}(\phi_{\alpha}^m\partial^{\mu}\phi_m^{\beta})
  (\phi_{\beta}^n\partial_{\mu}\phi_n^{\alpha})\\ \nonumber
  &&-\frac{1}{2}G^{mn}(\phi)
  \partial^{\mu}\phi_m^{\beta}\partial_{\mu}\phi_{\beta n}\;.
 \end{eqnarray}
Looking back to (\ref{dualaction}) one infers that this coincides
(after coupling to gravity) with the dimensionally reduced action
after dualisation. By construction, this $\sigma$-model action has
a non-linear rigid $SL(d+1,\mathbb{R})$ symmetry. The group action
on the fields corresponding to this enhanced symmetry will be
determined in the following.

To start with, we remind the reader that for a coset space $G/H$ the
$G$ acts rigidly on a group element like in (\ref{groupelement})
by left multiplication, while the maximal compact subgroup $H$ acts
by local right multiplication. Infinitesimally, it reads in the given case
 \bea\label{cosetaction}
  \delta{\cal V} = \hat{g}{\cal V}-{\cal V}\hat{h}(x)\;,
 \eea
where $\hat{g}\in sl(d+1,\mathbb{R})$ and $\hat{h}(x)\in so(d+1)$.
With (\ref{cosetaction}) one finds that the $sl(d,\mathbb{R})$
acts on the fields linearly as
 \bea
  \delta_{\lambda}(K)\phi_m^{\alpha}=\lambda^a_{\hspace{0.3em}b}
  (K_{\hspace{0.3em}a}^b)_m^{\hspace{0.3em}n}\phi_n^{\alpha}\;, \qquad
  \delta_{\lambda}(K)\varphi_m = \lambda^a_{\hspace{0.3em}b}
  (K_{\hspace{0.3em}a}^b)_m^{\hspace{0.3em}n}\varphi_n\;.
 \eea
In contrast, the symmetries generated by $e_a$ act non-linearly
as a shift,
 \bea
  \delta_{\lambda} (e_a)\phi_m^{\alpha} = 0\;, \qquad \delta_{\lambda}(e_a)\varphi_m = -\lambda_m\;,
 \eea
while the $\hat{e}$ transformations act in accordance with (\ref{liealgebra})
as rescalings,
 \bea
  \delta_{\lambda}(\hat{e})\phi_m^{\alpha} = \lambda\phi_m^{\alpha} \;, \qquad
  \delta_{\lambda}(\hat{e})\phi = d\lambda \phi\;, \qquad
  \delta_{\lambda}(\hat{e})\varphi_m = (d+1)\lambda\varphi_m\;.
 \eea
For the variations induced by the $f^a$ one has to take into
account that it is necessary to add a compensating local $SO(d+1)$
transformation in order to restore the triangular gauge in
(\ref{groupelement}). Choosing for this the transformation
parameter $\xi^{\alpha}=\phi\lambda^m\phi_m^{\alpha}$ (where
$\lambda^m$ parametrizes the rigid transformation), one finds
correspondingly the non-linear group action
 \bea
  \delta_{\lambda}(f_a)\phi_m^{\alpha} = (\lambda^l\varphi_m)\phi_l^{\alpha}\;, \qquad
  \delta_{\lambda}(f_a)\varphi_m = (\lambda^l\varphi_l)\varphi_m\;.
 \eea

\section{Explicit reduction without
truncation}\setcounter{equation}{0}\label{B} In this appendix we
compute the `dimensionally' reduced action, i.e.~in Yang-Mills
form, directly from the $D$-dimensional Einstein-Hilbert term
 \bea\label{hilbert}
  S_{\rm EH}=-\int d^Dx\hspace{0.2em}EE_A^M E_B^N
  (\partial_{M}\omega_{N}^{\hspace{0.5em}AB}-\partial_{M}\omega_{N}^{\hspace{0.5em}AB}+
  \omega_M^{\hspace{0.5em}AC}\omega_{N C}^{\hspace{1.1em}B}-
  \omega_N^{\hspace{0.5em}AC}\omega_{M C}^{\hspace{1.1em}B})\;.
 \eea
The spin connection in flat indices is defined in terms of the
coefficients of anholonomity as
 \bea\label{anholo}
   \omega_{ABC} =
   \frac{1}{2}\left(\Omega_{ABC}-\Omega_{BCA}+\Omega_{CAB}\right)\;,
   \qquad
   \Omega_{AB}^{\hspace{1.3em}C}=2E_{[A}^M E_{B]}^N\partial_M E_N^C\;.
 \eea
This gives rise to the following components
 \bea\label{holocom}
  \begin{split}
   \Omega_{abc}&=2\phi e^{\mu}_{[a}e^{\nu}_{b]}D_{\mu}e_{\nu c}
   +2\eta_{c[a}e_{b]}^{\nu}D_{\nu}\phi\;, \\
   \Omega_{ab\alpha}&=\phi^2F_{ab\hspace{0.2em}\alpha}
   :=\phi^2 e_a^{\mu}e_b^{\nu}F_{\mu\nu}^n\phi_{n\alpha}\;, \\
   \Omega_{\alpha bc}&=g\phi_{\alpha}^m e_b^{\nu}D_m e_{\nu c}\;,\\
   \Omega_{\alpha\beta\gamma}&=2g\phi_{[\alpha}^m\phi_{\beta]}^n
   \partial_m\phi_{n\gamma}\;, \\ 
   \Omega_{\alpha\beta c}&=0\;, \\
   \Omega_{a\alpha\beta}&= \phi\phi_{\alpha}^m e_a^{\mu}
   D_{\mu}\phi_{m\beta}\; .
  \end{split}
 \eea
Here we have introduced an internal covariant derivative
 \bea\label{Dm}
  D_m e_{\mu}^a = \partial_m e_{\mu}^a - (\phi^{-1}\partial_m\phi)e_{\mu}^a\;,
 \eea
which transforms covariantly under local $\hat{v}_d$ transformations
(albeit in a different representations as $e_{\mu}^a$):
 \bea
  \delta_{\xi}(D_m e_{\mu}^a) = g\xi^n\partial_n(D_m e_{\mu}^a)
  +g\partial_m\xi^n D_n e_{\mu}^a + g\partial_n\xi^n D_m e_{\mu}^a\;.
 \eea

Inserting (\ref{holocom}) into the Einstein-Hilbert action
(\ref{hilbert}) and dropping total derivatives results in
 \bea\label{EHYMfull}
  \begin{split}
   S_{\rm EH}=\int d^3x d^dy \hspace{0.2em}e\big[ &-R_3^{\rm cov}(e)
   -\frac{1}{4}\phi^2G_{mn}(\phi)F^{\mu\nu\hspace{0.2em}m}
   F_{\mu\nu}^{\hspace{0.1em}n}+\phi^{-2}D^{\mu}\phi D_{\mu}\phi \\
   &+ \frac{1}{2}(\phi_{\alpha}^mD^{\mu}\phi_m^{\gamma})
   (\phi_{\gamma}^nD_{\mu}\phi_n^{\alpha})-\frac{1}{2}G^{mn}(\phi)
   D^{\mu}\phi_m^{\beta}D_{\mu}\phi_{n\beta} \\
   &+g^2\phi^{-2}R(\phi)-\frac{1}{2}gF^{ab\hspace{0.2em}m}e_{[a}^{\nu}D_me_{\nu b]}\\
   &-\frac{1}{2}g^2\phi^{-2}G^{mn}(\phi)(e_a^{\nu}D_me_{\nu}^b)
   (e_b^{\mu}D_ne_{\mu}^a) \\
   &-\frac{1}{2}g^2\phi^{-2}G^{mn}(\phi)g^{\mu\nu} D_me_{\mu}^aD_n e_{\nu a} \\
   &-g^2\phi^{-2}G^{mn}(e^{a\mu}D_m e_{\mu a})(e^{b\nu}D_n e_{\nu b})
   \big]\;.
  \end{split}
 \eea
As claimed in the main text, the action appears in form which is
manifestly invariant under local $\hat{v}_d$ transformations.
Moreover, it contains the spin-2 mass terms and the scalar
potential given in (\ref{gaugeind}). Let us finally note that even
though the invariance of (\ref{EHYMfull}) under local
$SO(1,2)\times SO(d)$ transformations is not obvious, it can be
checked explicitly. In particular, it turns out that, quite
surprisingly, $R_3^{\rm cov}$ is not invariant under all local
$SO(1,2)$ transformations, but its variation cancels against the
variation of the term proportional to $F^{ab\hspace{0.1em}m}$.

\end{appendix}

\end{document}